\documentstyle[12pt,psfig]{article}

\newcommand{\resection}[1]{\setcounter{equation}{0}\section{#1}}

\def\to{\rightarrow}

\def\EQ{\begin{equation}}
\def\EN{\end{equation}}
\def\bea{\begin{eqnarray}}
\def\eea{\end{eqnarray}}

\begin{document}
\oddsidemargin 5mm
\setcounter{page}{0}
\newpage     
\setcounter{page}{0}
\begin{titlepage}
%\begin{flushright}
%hep-th/9712111 \\
%.../97/...
%\end{flushright}
\vspace{0.5cm}
\begin{center}
{\large {\bf Susceptibility amplitude ratios \\
in the two-dimensional Potts model and percolation}} \\
\vspace{1.5cm}
{\bf G.~Delfino$^{a}$, G.T.~Barkema$^{b}$ and John Cardy$^{c,d}$} \\
\vspace{0.8cm}
$^a${\em Laboratoire de Physique Th\'eorique et Hautes Energies\\
Universit\'e Pierre et Marie Curie, Tour 16 \mbox{1}$^{er}$ \'etage, 4
place Jussieu\\
75252 Paris cedex 05, France} \\
$^b${\em Theoretical physics, Utrecht University\\
Princetonplein 5, 3584 CC Utrecht, the Netherlands}\\
$^c${\em Theoretical Physics, University of Oxford\\
1 Keble Road, Oxford OX1 3NP, United Kingdom} \\ 
$^d${\em All Souls College, Oxford} \\
\end{center}
\vspace{6mm}
\begin{abstract}
\noindent
The high-temperature susceptibility of the $q$-state Potts model behaves as
$\Gamma|T-T_c|^{-\gamma}$ as $T\to T_c+$, while for $T\to T_c-$ one may
define both longitudinal and transverse susceptibilities, with the same
power law but different amplitudes $\Gamma_L$ and $\Gamma_T$. We extend
a previous analytic calculation of the universal ratio $\Gamma/\Gamma_L$
in two dimensions to the low-temperature ratio $\Gamma_T/\Gamma_L$, and
test both predictions with Monte Carlo simulations for $q=3$ and 4.
The data for $q=4$ are inconclusive owing to large corrections to
scaling, while for $q=3$ they appear consistent with the prediction for
$\Gamma_T/\Gamma_L$, but not with that for $\Gamma/\Gamma_L$. A simple
extrapolation of our analytic results to $q\to1$ indicates a similar
discrepancy with the corresponding measured quantities in percolation.
We point out that stronger assumptions were made in the
derivation of the ratio $\Gamma/\Gamma_L$, and our work suggests that
these may be unjustified.
\end{abstract}
\vspace{5mm}
\end{titlepage}
\newpage

\setcounter{footnote}{0}
\renewcommand{\thefootnote}{\arabic{footnote}}

\newpage
\resection{Introduction}
Two statistical systems can be characterised by the same internal symmetries
and still differ in their microscopic realisation. This difference will be
observable as far as the correlation length is not much larger than the 
microscopic length scale. Nearby a second order phase transition point,
however, the microscopic details become irrelevant and the two systems appear 
as representatives of the same {\em universality class}. 

In the characterisation of universal behaviour, which is one of the basic
tasks of statistical mechanics, one can distinguish different steps. The 
first one is the determination of the universal features of the critical point
(first of all the critical exponents). In two dimensions this goal was 
achieved with the solution of conformal field theories \cite{BPZ,ISZ}. The 
second natural step is the study of the scaling region surrounding the fixed
point. The leading behaviour of a physical quantity in this region is 
assigned in terms of a critical amplitude multiplying the suitable power
of the temperature. The critical amplitudes depend on metric factors, but
they can be used to construct universal combinations which characterise the 
scaling region \cite{PHA}. 

It is clear that  the computation of the universal 
amplitude combinations requires a solution of the theory away from criticality.
This has become possible over the last years for a large class of 
two-dimensional quantum field theories characterised by the presence of an 
infinite number of integrals of motion (integrable field theories) 
\cite{Taniguchi}. They describe the scaling limit of the isotropic statistical 
models which are solved on the lattice, but also of many
others whose lattice solution is not available. In particular, the universal
amplitude ratios for such a basic model as the Ising model in a magnetic 
field have been computed {\em exactly} in this framework \cite{ratios}. 
More generally, integrable field theory provides accurate approximations for 
the amplitude ratios \cite{CM,DC}. 

For the purpose of comparison with the results provided by integrable field 
theory, it is clearly desirable to obtain independent estimates for the 
universal quantities. In view of the difficulties of other traditional 
approaches (accurate series expansions are available only for few models, and 
$d=2$ is normally too far from the upper critical dimension to obtain 
reliable estimates through the $\epsilon$-expansion), numerical simulations 
appear as a most valuable source of data. The field theoretical predictions 
concerning the scaling limit of an infinite system can in principle be tested
numerically working in a range of temperature for which the correlation 
length is much larger than the lattice spacing and much smaller than the 
lattice size. In practice, however, both the location of this temperature
window and the required lattice sizes are model dependent and not obvious to
identify.

This paper deals with universal amplitude ratios for the two-dimensional 
$q$-state Potts model and the related problem of isotropic percolation.
These have been the subject of a general study in the framework of integrable
field theory in Ref.\,\cite{DC}. Here we mainly focalise on the susceptibility
amplitude ratios presenting new theoretical predictions for the ratio (not 
considered in \cite{DC}) of the transverse and longitudinal susceptibilities 
below $T_c$, and a Monte Carlo study for $q=3$ and 4.

The $q$-state Potts model \cite{Potts,Wu} is defined by the lattice Hamiltonian
\EQ
H=-J\sum_{(x,y)}\delta_{s(x),s(y)}\,,
\EN
where the sum is over nearest neighbours and the site variable $s(x)$ can 
assume $q$ possible values (colours). The model is clearly invariant under the
group of permutations of the colours. In the ferromagnetic case $J>0$ we 
are interested in, the states in which all the sites have the same colour 
minimise the energy and the system exhibits spontaneous magnetisation at 
sufficiently low temperatures. There exists a critical temperature $T_c$ above 
which the thermal fluctuations become dominant and the system is in a 
disordered phase. We will consider the Potts model in two dimensions in the
range of the parameter $q$ for which the phase transition at $T=T_c$ is 
continuous, namely $q\leq 4$ \cite{Baxter}.

Let us introduce the spin variables
\EQ
\sigma_i(x)=\delta_{s(x),i}-\frac{1}{q}\,,\hspace{1cm}i=1,2,\ldots,q
\label{sigma}
\EN
satisfying the condition
\EQ
\sum_{i=1}^q\sigma_i(x)=0\,\,.
\label{constraint}
\EN
When $T>T_c$ all values of the site variable occur with equal probability
$1/q$. At low-temperature, however, one of the $q$ degenerate ground states is
selected out by spontaneous symmetry breaking. This might be done either
by imposing a symmetry-breaking field which is allowed to tend to zero
after taking the thermodynamic limit, or by imposing symmetry-breaking
boundary conditions before taking the limit. Without loss of generality
we choose the colour of the selected ground state at $T<T_c$ to correspond 
to $i=1$.
Then, for any temperature, we can write
\EQ
\langle\sigma_i\rangle=\frac{q\delta_{i1}-1}
{q-1}\,M\,,
\label{M}
\EN
where $M$ denotes the `longitudinal' spontaneous magnetisation 
$\langle\sigma_1\rangle$ and vanishes at $T>T_c$.

The connected spin--spin correlation functions are given by
\EQ
G_{ij}(x)=\langle\sigma_i(x)\sigma_j(0)\rangle-\langle\sigma_i\rangle
\langle\sigma_j\rangle\,\,.
\label{connected}
\EN
If $\nu_i$ denotes the fraction of sites with 
colour $i$, the magnetic susceptibilities per site can be written as
\EQ
\chi_i=\sum_x G_{ii}(x)=\langle \nu_i^2\rangle-\langle\nu_i\rangle^2\,\,.
\label{chii}
\EN
Of course, $\chi_i=\chi$ at $T>T_c$, while in the low-temperature phase we 
have to distinguish between the longitudinal susceptibility $\chi_L=\chi_1$
and the transverse susceptibility $\chi_T=\chi_{i\neq 1}$.
In the vicinity of the critical point, for $q<4$, the susceptibilities behave as
\EQ
\chi_i\simeq\Gamma_i\,t^{-\gamma}\,\,,
\EN
where $t=|T-T_c|/T_c$.
Denoting $\Gamma$, $\Gamma_L$ and $\Gamma_T$ the critical amplitudes 
associated respectively to $\chi$, $\chi_L$ and $\chi_T$, we have the 
two universal amplitude ratios 
\EQ
\Gamma/\Gamma_L\,\,,\hspace{1cm}\Gamma_T/\Gamma_L\,\,.
\label{ratios}
\EN
For $q=4$ it is well known that quantities like the susceptibility
asymptotically exhibit multiplicative logarithmic correction factors of
the form $|\ln|t||^{\bar\gamma}$ \cite{CNS,SS}.  These are due to a
marginally irrelevant operator.  The analytic calculation of
Ref.~\cite{DC} was performed in the continuum massive field theory
corresponding to a point on the outflowing renormalisation group
trajectory.  From this point of view, the logarithmic factors arise
only when the parameters of the continuum theory are expressed in terms
of those of the bare theory. We therefore expect that predictions for
such universal quantities as the amplitude ratios above still to be
valid when applied to ratios in which the leading logarithmic factors
cancel.

In the next section we recall the link between the isotropic percolation 
problem and the $q\rightarrow 1$ limit of the Potts model, and show how in 
this limit the ratios (\ref{ratios}) provide some universal information about 
percolation clusters. In section 3 we recall the origin of the theoretical 
predictions and present the new analytic results
for the ratio $\Gamma_T/\Gamma_L$ and its percolation analogue. Section 4 is
devoted to a Monte Carlo study of the ratios (\ref{ratios}) in the 
Potts model for $q=3$ and 4 before discussing the theoretical and numerical 
results in the final section.

\resection{Connection with percolation}
Percolation is the geometrical problem in which bonds
%\footnote{We refer here to {\em bond} percolation; in {\em site} 
%percolation, sites rather than bonds are occupied with probability $p$. Of 
%course, all universal results are independent of this distinction.}  
are randomly distributed on a lattice with occupation probability $p$ 
\cite{SA}. 
A set of bonds forming a connected path on the lattice is called a {\em 
cluster}. There exist a critical value $p_c$ of the occupation probability 
above which an infinite cluster appears in the system; $p_c$ is called the 
{\em percolation threshold}. If $N$ is the total number of bonds in the 
lattice, the probability of a configuration with $N_b$ occupied bonds is 
$p^{N_b}(1-p)^{N-N_b}$. Hence, the average of a quantity $X$ over all 
configurations ${\cal G}$ is 
\EQ
\langle X\rangle=\sum_{\cal G}X\,p^{N_b}(1-p)^{N-N_b}\,\,.
\label{average}
\EN

It is well known that the percolation problem can be mapped onto the limit 
$q\rightarrow 1$ of the $q$-state Potts model \cite{KF}. In fact, if we define
$z=e^{J/T}-1$, the partition function of the Potts model can be written in the 
form
\EQ
Z=\mbox{Tr}_s\prod_{(x,y)}(1+z\delta_{s(x),s(y)})\,\,.
\label{partition}
\EN
A graph ${\cal G}$ on the lattice can be associated to each Potts configuration
by drawing a bond between two sites with the same colour. In the above 
expression, a power of $z$ is associated to each bond in the graph. Taking 
into account the summation over colours one arrives to the expansion
\cite{Baxter2}
\EQ
Z=\sum_{\cal G}q^{N_c}z^{N_b}\,,
\label{newpartition}
\EN
where $N_b$ is the total number of bonds in the graph ${\cal G}$ and $N_c$
is the number of clusters in ${\cal G}$ (each isolated
site is also counted as a cluster). In terms of the partition function
(\ref{newpartition}) the $q$-state Potts model is well defined even for 
noninteger values of $q$. The average of a quantity $X$ can be written as
\EQ
\langle X\rangle_q=Z^{-1}\sum_{\cal G}X\,q^{N_c}z^{N_b}\,\,.
\EN
Hence, it is sufficient to make the formal identification $z=p/(1-p)$ to see
that $\langle X\rangle_1$ coincides with the percolation average 
(\ref{average}). For $q\neq 1$ the Potts model describes a generalised 
percolation problem in which each cluster can assume $q$ different colours.
The presence of a spontaneous magnetisation $M$ at $T<T_c$ reflects the 
appearance of an infinite cluster at $p>p_c$. 

Let $P$ denote the probability that a site belongs to the infinite cluster
($P=0$ for $p<p_c$). Then, for any value of $p$, the probability that the site
$x$ has colour $k$ is
\EQ
\langle\delta_{s(x),k}\rangle=P\delta_{k1}+\frac{1}{q}\,(1-P)\,\,.
\EN
Recalling Eqs.\,(\ref{sigma}) and (\ref{M}), we obtain
\EQ
P=\frac{q}{q-1}\,M\,\,.
\EN
Consider now two sites located at $x$ and $y$, and call $P_i$ the probability
that they are both in the infinite cluster, $P_f$ the probability that they
are in the same finite cluster, $P_{ff}$ the probability that they are in
different finite clusters, and $P_{if}$ the probability that $x$ is in the 
infinite cluster while $y$ is in a finite one. The probability that $x$ has 
colour $k$ and $y$ has colour $j$ can be expressed as
\EQ
\langle\delta_{s(x),k}\delta_{s(y),j}\rangle=P_i\delta_{k1}\delta_{j1}+
P_f\frac{1}{q}\,\delta_{kj}+P_{ff}\frac{1}{q^2}+P_{if}\frac{1}{q}
(\delta_{k1}+\delta_{j1})\,\,.
\EN
Since $P_i+P_f+P_{ff}+2P_{if}=1$ and $P_i+P_{if}=P$, the two-point correlations
only depend on two independent functions of $x-y$, say $P_i$ and $P_f$. 
For the connected spin--spin correlators (\ref{connected}) one finds
\bea
G_{kj}(x) & = & \left(\delta_{k1}\delta_{j1}-\frac{1}{q}\,(\delta_{k1}+
\delta_{j1})+\frac{1}{q^2}\right)P_i(x)+\left(\frac{1}{q}\,\delta_{kj}-
\frac{1}{q^2}\right)P_f(x)\nonumber\\
          & - & \left(\delta_{k1}-\frac{1}{q}\right)\left(\delta_{j1}-
\frac{1}{q}\right)P^2\,\,.
\label{corr}
\eea

Restrict from now on our attention to the case of ordinary percolation, 
so that the limit $q\rightarrow 1$ is understood in all the subsequent 
equations. From the previous equation we obtain
\bea
&& G_{11}(x)=(q-1)\,P_f(x)\,,\\
&& G_{kk}(x)=P_i(x)-P^2\,\,,\hspace{1cm}k\neq 1\,\,.
\eea
The average size of {\em finite} clusters is given by
\EQ
S=\sum_x P_f(x)=\frac{1}{q-1}\sum_x G_{11}(x)\,\,.
\label{size}
\EN
Nearby the percolation threshold this quantity behaves as 
\EQ
S\simeq \sigma_\pm|p_c-p|^{-\gamma}\,\,,
\EN
where the subscripts $+$ and $-$ refer to $p<p_c$ and $p>p_c$, respectively,
and $\gamma$ is the Potts critical exponent evaluated at $q=1$. 
Equation (\ref{size}) implies
\EQ
\frac{\sigma_+}{\sigma_-}=\frac{\Gamma}{\Gamma_L}\,\,.
\EN
The quantity
\EQ
S'=\sum_x(P_i(x)-P^2)=\sum_x G_{kk}(x)\,,\hspace{1cm}k\neq 1
\EN
is a measure of the short range correlations inside the infinite cluster and
behaves near criticality as
\EQ
S'\simeq \sigma'|p_c-p|^{-\gamma}\,\,.
\EN
One can then introduce a second universal ratio $\sigma'/\sigma_-$ whose 
relation with the Potts susceptibility amplitudes is
\EQ
\frac{\sigma'}{\sigma_-}=(q-1)\,\frac{\Gamma_T}{\Gamma_L}\,\,.
\label{lt}
\EN

\resection{Analytic results}
The scaling limit of the $q$-state Potts model is an integrable field theory
\cite{DF,Taniguchi}, 
and this fact allows the evaluation of the correlation functions through the 
form factor approach, which is of general applicability within integrable 
field theory. This programme was carried out for the $q$-state Potts model 
in Ref.\,\cite{DC}. We just recall here the basic steps of the procedure,
referring the reader to that paper for all the details. 

The starting point is the exact scattering description of the low-temperature
phase of the model determined by Chim and Zamolodchikov \cite{CZ}. Since 
at $T<T_c$ the model exhibits $q$ degenerate vacua, the elementary excitations
entering this scattering description are kinks interpolating among the 
different vacua. The knowledge of the $S$-matrix allows the computation of
the matrix elements (form factors) $\langle 0|\Phi(0)|n\rangle$ entering the
spectral decomposition of the correlation functions:
\EQ
\langle\Phi_1(x)\Phi_2(0)\rangle=\sum_{n=0}^{\infty}
\langle 0|\Phi_1(0)|n\rangle\langle n|\Phi_2(0)|0\rangle e^{-|x|E_n}\,,
\label{spectral}
\EN
where $E_n$ denotes the total energy of the $n$-particle state $|n\rangle$.
It is known that in integrable models the spectral series (\ref{spectral}) 
exhibit remarkable convergence properties, and that, in particular, very
accurate estimates of integrated correlators can be obtained retaining only
the terms of the series containing no more than two particles (two-particle
approximation). In Ref.\,\cite{DC}, the one- and two-particle form factors of 
the energy, spin and disorder operators were computed in both phases of the 
model, the information about the high-temperature phase being obtained by 
duality. The two-particle approximation for the correlators was then used
to evaluate a series of universal amplitude ratios, including 
$\Gamma/\Gamma_L$. Ref.\,\cite{DC} also contains all
the necessary information for the computation (within the same approximation) 
of the ratio $\Gamma_T/\Gamma_L$, which however was not discussed in that 
paper. We give in Table~1 the results corresponding to $q=2,3,4$.

Due to technical difficulties, the form factor equations for the spin operator
could be solved only for $q=2,3,4$ in Ref.\,\cite{DC}, a limitation which 
prevents the analytic continuation to the percolation point $q=1$ for those
amplitude ratios which are related to correlation functions of the spin 
operator. An estimate of these percolation ratios ($\sigma_+/\sigma_-$, in
particular) was however proposed in terms of a simple (quadratic) extrapolation
to $q=1$ of the results obtained for $q=2,3,4$. We do here the same thing 
for the low-temperature ratio $\sigma'/\sigma_-$. Using the values of 
$\Gamma_T/\Gamma_L$ given in Table~1 for the extrapolation of 
Eq.\,(\ref{lt}), and explicitly incorporating the factor of $(q-1)$,
we find\footnote{This is the result of the extrapolation 
performed in the variable $\lambda$ (related to $q$ by 
$\sqrt{q}=2\sin(\pi\lambda/3)$) in which all the results originating from the 
scattering theory are analytic. Extrapolating in $q$ gives 
$\sigma'/\sigma_-\approx 1.43$.}
\EQ
\frac{\sigma'}{\sigma_-}\approx 1.49\,\,.
\label{149}
\EN
According to the considerations developed in \cite{DC}, we estimate an accuracy
of order 1\% for our predictions of $\Gamma_T/\Gamma_L$, while we allow for a 
10\% error on the value of $\sigma'/\sigma_-$ to take into account the 
uncertainty coming from the extrapolation procedure.

We conclude this section with a reminder of the theoretical predictions for 
the correlation length amplitude ratio which will also be considered in the 
next section. The ``true'' correlation length $\xi$ is defined through the
large distance decay of the spin-spin correlation function
\EQ
\langle\sigma_i(x)\sigma_i(0)\rangle\sim e^{-|x|/\xi}\,\,,
\EN
and is determined as the inverse of the total mass of the lightest state
entering the spectral series (\ref{spectral}). For the spin operator at 
$q\leq 3$, this lightest state is the one-kink state for $T>T_c$, and the 
two-kink state for $T<T_c$. For $q>3$ one has to take into account that the 
two-kink state gives rise to a bound state whose mass equals $\sqrt{3}\,m$ at
$q=4$, $m$ being the mass of the kink. Hence, denoting by $\xi_\pm$ the 
critical amplitudes of the ``true'' correlation length in the two different
phases, the exact results for the ratio for integer $q$
are those given in Table~1. We also include our results for $\Gamma/\Gamma_L$ 
taken from
Ref.~\cite{DC}.

\begin{table}
\caption{Analytic results for susceptibility and correlation length
ratios. Those for the susceptibility are in the two-particle
approximation, while those for the correlation length are exact.}
\vspace{0.5cm}
\centerline{
\begin{tabular}{|c|c|c|c|}\hline
$ q $ & $2$        & $3$        & $4$      \\ \hline
$\Gamma_T/\Gamma_L$ & $ 1 $ & $ 0.327 $  & $ 0.129 $  \\
$\Gamma/\Gamma_L$ & $ 37.699 $ & $ 13.848 $ & $ 4.013 $  \\
$\xi_+/\xi_-$ & $ 2 $ & $ 2 $  &  $\sqrt{3}$ \\ \hline
\end{tabular}}
\end{table}

\resection{Computer simulations}

Computer simulations are performed on the three- and four-states Potts
model on an $L \times L$ square lattice with periodic (helical)
boundary conditions. The Wolff single-spin cluster algorithm
Wolff~\cite{wolff} was used, implemented as described in
Ref.~\cite{nbbook}. A random configuration is generated, and
thermalised by applying cluster moves until each spin is statistically
flipped 500 times for the three-states Potts model, 1000 times for the
four-state Potts model with $L\leq400$, or 2000 times for the
four-state Potts model with larger lattices.  After the thermalisation,
a sequence of 1000 (or for the smallest lattice size 2000)
configurations is generated, in which consecutive configurations are
separated by a sequence of cluster moves in which each spin is
statistically visited 10 times. In each configuration, the
magnetisation $\nu_i$, the fraction of spins in each state $i$, is
determined. From these numbers, the magnetic susceptibility above the
critical temperature is calculated using Eq.~(\ref{chii}) and averaging
over $i$.  Below the critical temperature, we expect to find
spontaneous symmetry breaking, but of course this cannot occur in a
finite system. One way to implement this would be to choose open
boundaries and to fix the spins on the boundary to a preferred state.
However, this brings in boundary effects and would necessitate using
much larger systems. Instead, we simply observe that in any given
configuration one colour of spin dominates, and in the thermodynamic
limit this will be the preferred orientation of the ground state.  Thus
we estimate the longitudinal susceptibility from the fluctuations in
the fraction of spins which are in the majority state, averaged over
all configurations, and the transverse susceptibility from the
fluctuations of the fraction of spins in each minority state, averaged
over the $(q-1)$ minority states in each configuration, and then over
all configurations. We expect that the difference between the result of
this method of estimation and that using fixed boundary conditions to
be of order $e^{-L/\xi}$, and thus exponentially suppressed in the
region we study\footnote{It should be noted that this is no longer the
case as $T\to T_c-$ at fixed $L$, and indeed we found that our
susceptibility ratios as estimated this way did not converge to unity
as the true ones must in this limit.}.

\begin{figure}
\psfig{figure=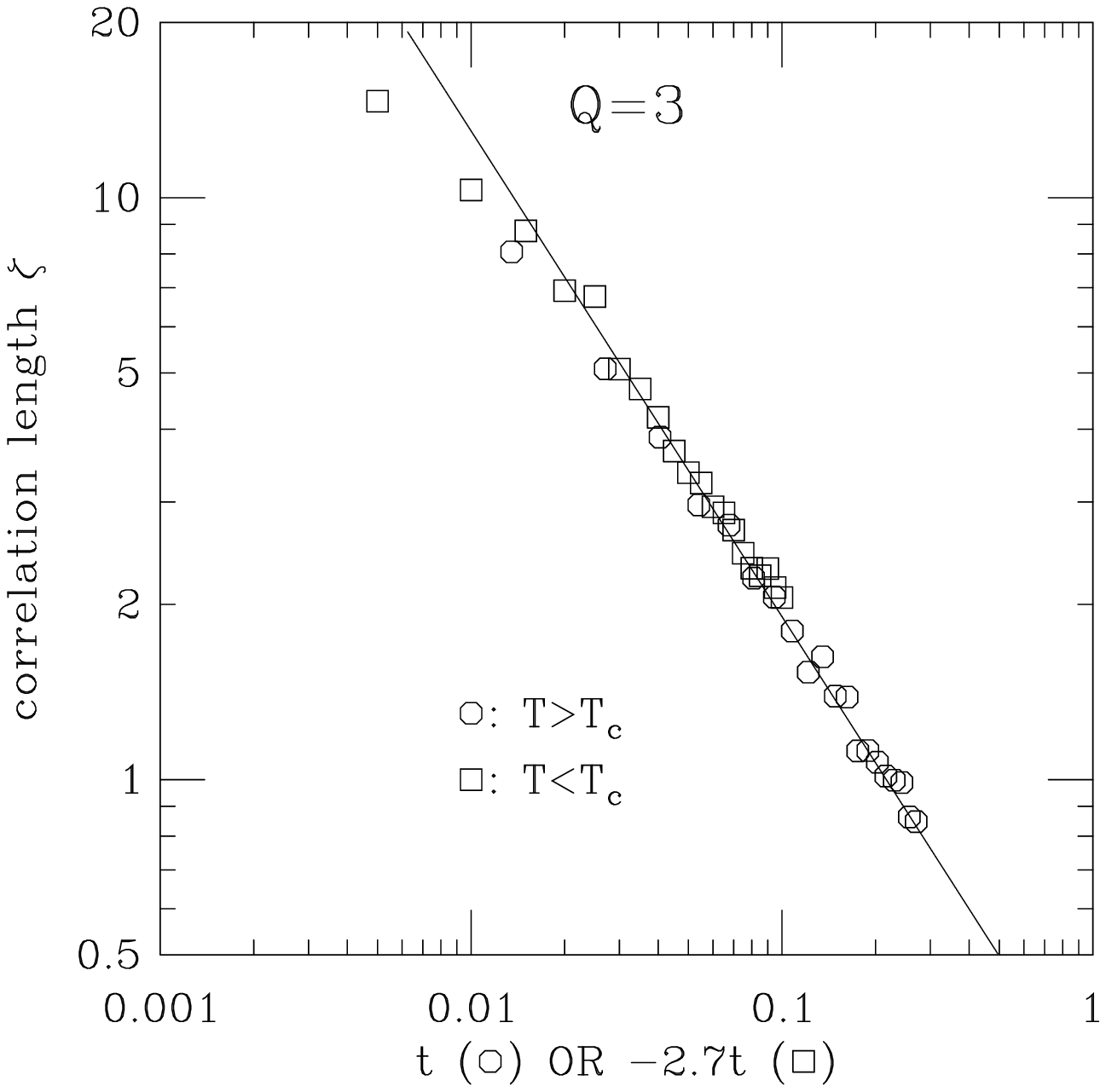,width=8cm}
\psfig{figure=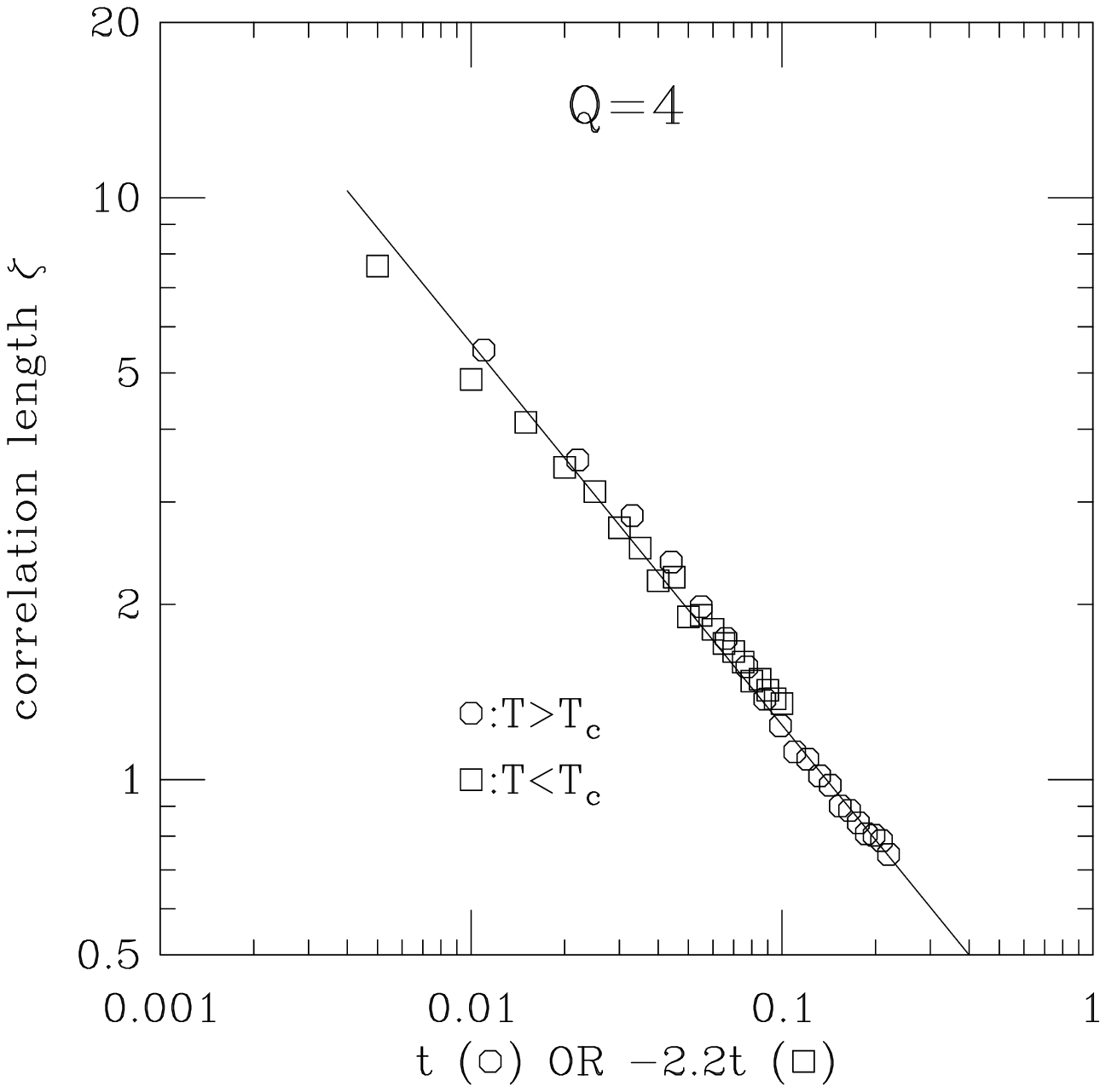,width=8cm}
\caption{Correlation length as a function of reduced temperature $t$
above the critical temperature (circles), and as a function of the
scaled reduced temperature $t'=-ct$ (squares), for the three-states
(top figure) and four-states Potts model (lower figure). The data
points overlap for $c_{Q=3}=2.7 \pm 0.5$ and $c_{Q=4}=2.2 \pm 0.3$.
The dashed lines show the expected behaviour $\xi \sim |t|^{-\nu}$.
\label{shifttemp}}
\end{figure}

We are interested in the ratios $\Gamma_T/\Gamma_L$ and $\Gamma/\Gamma_L$,
as discussed in the previous sections.  The measurements should be
performed at temperatures where the correlation length $\xi$ is large
compared to the lattice spacing, but small compared to the system
size.  To determine approximately the middle of the appropriate
temperature regime, we first measured the spin-spin correlation
function above and below the critical temperature, in simulations of
the $200 \times 200$ three- and four-states Potts models.  In figure
\ref{shifttemp} we plot the correlation length as a function of reduced
temperature $t=(T-T_c)/T_c$ above the critical temperature, and on top
of that the correlation length as a function of the scaled reduced
temperature $t'=-c(T-T_c)/T_c$ below the critical temperature. For the
three-state Potts model, the curves collapse for $c_{Q=3}=2.7 \pm 0.5$,
for the four-state Potts model for $c_{Q=4}=2.2 \pm 0.3$; both values
are in agreement with the theoretical expectations
$c_{Q=3}=2^{1/\nu}=2.297$ and $c_{Q=4}=(\sqrt{3})^{1/\nu}=2.280$.

\begin{figure}
\psfig{figure=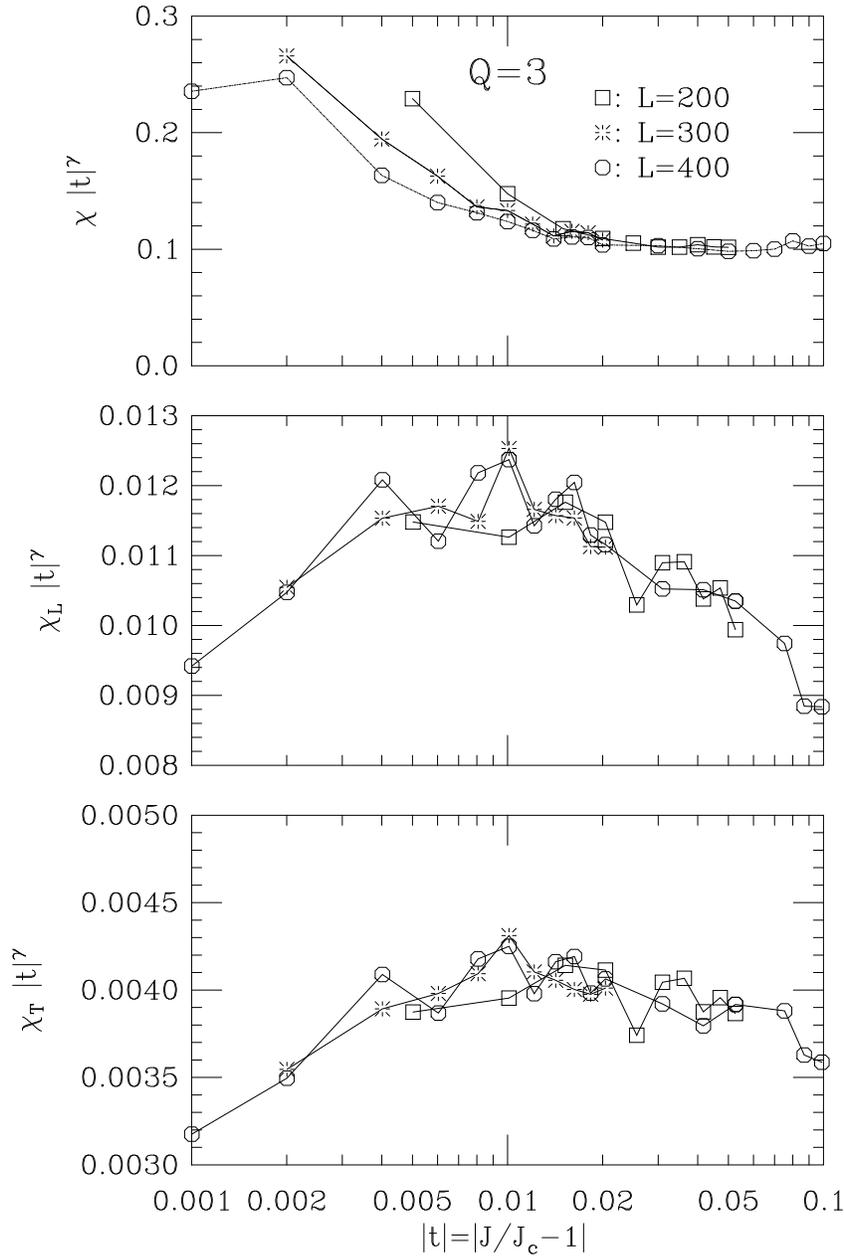,width=8cm}
\caption{Susceptibilities for the 3-state Potts model, above and below
$T_c$, rescaled by the expected asymptotic power-law dependence
$|t|^{-\gamma}$.  These plots exhibit the evidence for a plateau region
in which both finite-size effects and corrections to scaling are small.
The height of the plateau then indicates the relevant amplitude
$\Gamma_i$.
\label{fig2}}
\end{figure}

For the three- and four-states Potts models, we measured the
susceptibility $\chi$ for lattice sizes $L=200$ to 1200, at
temperatures $t_+$ above $T_c$.  We found surprising difficulty in
identifying a window where both finite-size effects and corrections to
scaling may simultaneously be ignored. The former are negligible in the
region for sufficiently large $t_+$ when the data for different,
sufficiently large, values of $L$ all collapse. The latter are
negligible if we observe a plateau in the collapsed data when it is
multiplied by $(t_+)^\gamma$.  An example of such a plateau is shown in
Fig.~\ref{fig2} for $q=3$.  We also found it more difficult to
identify the scaling window for $T>T_c$ than below the critical
temperature. This may be for two reasons: first the correlation length
amplitudes are larger above $T_c$, so that one needs to go to larger
values of $|t|$ to get rid of finite size effects; and second, periodic
boundary conditions move the peak in the finite-size susceptibility to
higher temperatures, pushing away the plateau in $\chi(t_+)^\gamma$.

After these initial difficulties we decided to repeat the exercise for
the $q=2$ Ising model and found precisely the same effect. Examination
of all three cases led us to the following prescription: we measured
the susceptibility above $T_c$ around the temperature where the
correlation length is around $\xi=\sqrt{L/2}$; we next measured the
longitudinal and transverse susceptibilities $\chi_l$ and $\chi_t$ at
the corresponding temperatures $t_-=-t_+/c$ below $T_c$ where the
correlation length is the same; the exact temperatures used in the
simulations are listed in table \ref{temptable}. From $\chi(t_+),
\chi_L(t_-)$ and $\chi_T(t_-)$ we obtain the required ratios using
\begin{eqnarray}
\frac{\Gamma_T}{\Gamma_L} & = & \frac{\chi_T(t_-)}{\chi_L(t_-)} \\
\frac{\Gamma  }{\Gamma_L} & = & \frac{\chi  (t_+)}{\chi_L(t_-)}
                                \left( \frac{t_+}{t_-} \right)^\gamma
\end{eqnarray}
where $\gamma_{Q=3}=13/9$ and $\gamma_{Q=4}=7/6$. The results are
presented in table \ref{temptable}. The statistical errors are two
standard deviations wide; they are obtained by repeating the same
procedure five or ten times, with different random number generator
seeds.

\begin{table}
\caption{Numerical estimates for the ratios $\Gamma_T/\Gamma_L$
and $\Gamma/\Gamma_L$.}
\vspace{0.5cm}
\centerline{
\begin{tabular}{|c|c|c|c|c|c|}\hline
$Q$ & $L$ & $t_+ $ & $t_-=-t_+/c$ & $\Gamma_T/\Gamma_L$ & $\Gamma/\Gamma_L$\\
\hline
3 & 200 & 0.006  & -0.002612 & 0.334 (0.004) & 19.6 (1.6) \\
  &     & 0.012  & -0.005224 & 0.336 (0.002) & 11.8 (0.6) \\
  &     & 0.024  & -0.01045  & 0.347 (0.006) &  9.5 (0.5) \\
  & 400 & 0.0042 & -0.001850 & 0.333 (0.005) & 14.7 (1.2) \\
  &     & 0.0085 & -0.003700 & 0.333 (0.006) & 10.4 (0.8) \\
  &     & 0.017  & -0.007401 & 0.338 (0.004) &  9.3 (0.7) \\
  & 800 & 0.003  & -0.001306 & 0.332 (0.004) & 11.4 (1.4) \\
  &     & 0.006  & -0.002612 & 0.333 (0.007) &  9.7 (0.4) \\
  &     & 0.012  & -0.005224 & 0.336 (0.005) &  9.6 (0.4) \\
\hline
4 & 200 & 0.0025 & -0.001096 & 0.153 (0.004) & 3.8 (0.4) \\
  &     & 0.005  & -0.002193 & 0.160 (0.002) & 3.0 (0.3) \\
  &     & 0.01   & -0.004386 & 0.166 (0.003) & 3.0 (0.2) \\
  & 400 & 0.0015 & -0.000658 & 0.151 (0.004) & 2.5 (0.6) \\
  &     & 0.003  & -0.001316 & 0.157 (0.003) & 2.7 (0.5) \\
  &     & 0.006  & -0.002632 & 0.163 (0.003) & 2.8 (0.2) \\
  & 800 & 0.0009 & -0.000395 & 0.146 (0.004) & 2.1 (0.4) \\
  &     & 0.0018 & -0.000789 & 0.152 (0.003) & 2.1 (0.3) \\
  &     & 0.0036 & -0.001579 & 0.159 (0.004) & 2.4 (0.2) \\
  &1200 & 0.0005 & -0.000219 & 0.153 (0.005) & 3.0 (0.7) \\
  &     & 0.001  & -0.000439 & 0.146 (0.005) & 1.9 (0.6) \\
  &     & 0.002  & -0.000877 & 0.154 (0.006) & 2.1 (0.4) \\
\hline
\end{tabular}}
\label{temptable}
\end{table}

\resection{Analysis and comparison with other work}

We first discuss the comparison of our numerical results in Table~2 with
the analytic predictions presented in Table~1. The most stable results
are those for the ratio $\Gamma_T/\Gamma_L$ for $q=3$. The analytic
prediction of 0.327 is just outside the statistical error bars, but we
note that for a fixed temperature there is a consistent trend towards
this value with increasing $L$. We conclude that the data support the
analytic prediction in this case, particularly given the small but
unknown errors of the two-particle truncation, which might be expected
to lie in the third decimal place. However, for the ratio
$\Gamma/\Gamma_L$, while the data is less stable, there is clear trend
towards a value below 10.0, with error bars which, although large,
appear to exclude the analytic prediction of 13.8.

\begin{figure}
\psfig{figure=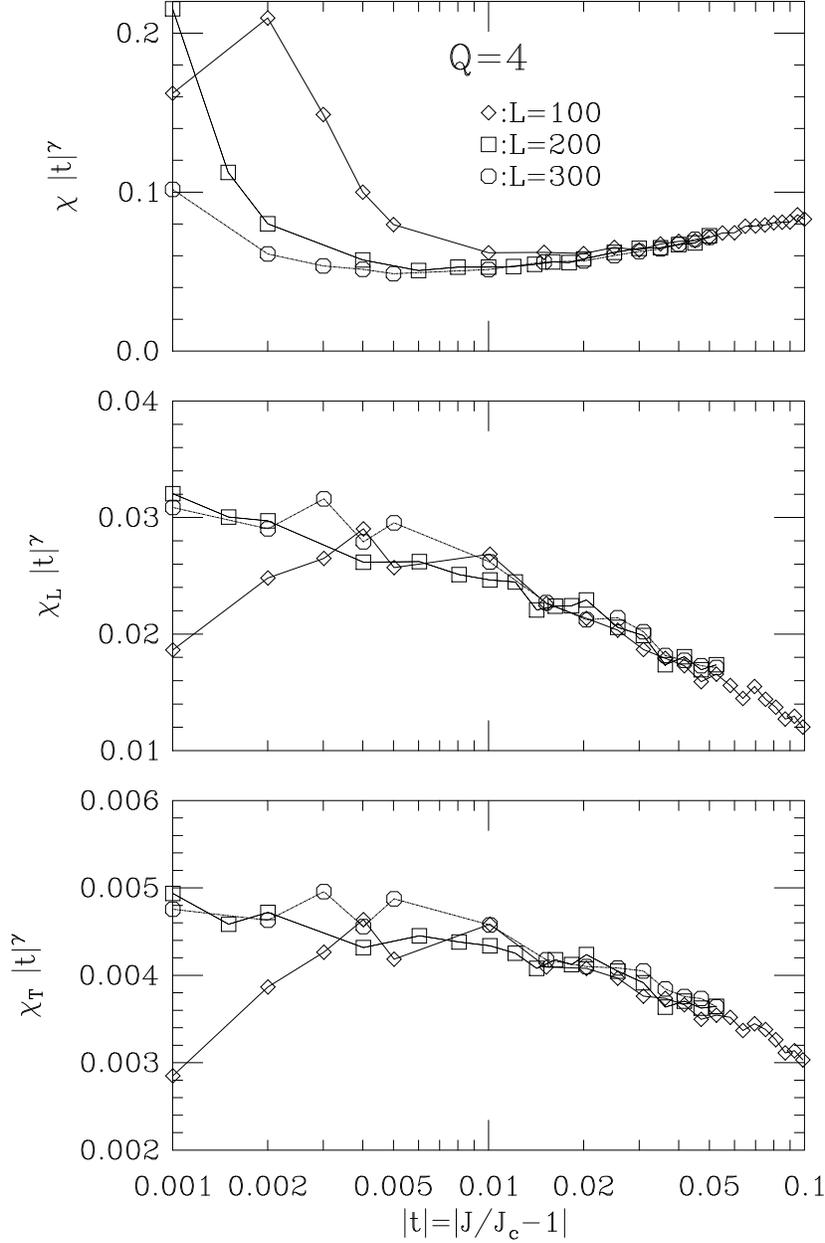,width=8cm}
\caption{Susceptibilities for the 4-state Potts model, above and below
$T_c$, rescaled by the expected asymptotic power-law dependence
$|t|^{-\gamma}$.  These plots show that although there are windows in
which finite-size effects and power-law corrections to scaling may be
ignored, the corresponding `plateaux' are in fact sloping. This is
presumably due to logarithmic corrections. Notice, however, that while
the data for $T<T_c$ are consistent with a multiplicative correction
$|\ln|t||^{\bar\gamma}$, with a positive $\bar\gamma$ as theoretically
expected, those for $T>T_c$ exhibit the opposite slope.
\label{fig3}}
\end{figure}

The situation for $q=4$ is more complex. The data for
$\Gamma_T/\Gamma_L$ appear fairly stable, yet even in the most
favourable case lie at least three standard deviations above the
analytic prediction. The situation for $\Gamma/\Gamma_L$ is even worse
as the results do not appear to be stable. As remarked earlier, in the
amplitude ratios the leading multiplicative logarithmic prefactors
should cancel, and even some of the non-leading terms \cite{SS}, but
there is no reason to suppose this is true for the $O(1/\ln|t|)$
corrections and further.  For $\Gamma_T/\Gamma_L$ it is conceivable
that these corrections are responsible for the discrepancy with the
analytic prediction, but we have not attempted to perform a fit
including the $O(1/\ln|t|)$ corrections, since at the reduced
temperatures we are working the neglect of the further corrections
cannot be justified.  The instability of the results for
$\Gamma/\Gamma_L$ may be understood on plotting the scaled
susceptibilities $\chi(t_+)^\gamma$ and $\chi_L(t_-)^\gamma$, which
should, asymptotically, reveal the logarithmic prefactors
$(\ln|t|)^{\bar\gamma}$, with $\bar\gamma=\frac34$. In fact (see
Fig.~\ref{fig3}) these have \em opposite \em slopes, indicating that,
in this region, the effective exponents $\bar\gamma$ have different
signs.  Once again, this is probably explained by the importance of
non-leading and non-universal $1/\ln|t|$ corrections. We deduce that
our numerical results are inconclusive for $q=4$.

We now compare our results with those of some other recent studies.

Salas and Sokal\cite{SS} made a detailed study of logarithmic
corrections in the $q=4$ model, including the susceptibility for
$T>T_c$. Our raw data appears to be consistent with theirs, in the
ranges of $t_+$ and $L$ for which they overlap. However, their main
goal was to extract the exponent $\bar\gamma$ of the leading
multiplicative logarithmic prefactor. They found that there was no
region in which they could isolate such a prefactor and eliminate
finite-size corrections.  In fact, in order to determine $\bar\gamma$
they had to take such corrections systematically into account using a
modified form of finite-size scaling. It is therefore no surprise that
it should be impossible to determine the asymptotic amplitude of such a
term from data taken over similar ranges.

Caselle et al. \cite{caselle} have also performed Monte Carlo
simulations of the $q=4$ model with a view to extracting the
susceptibility amplitude ratios and also that involving the
magnetisation. Before taking into account any logarithmic corrections,
these authors' estimates for the ratio $\Gamma/\Gamma_L$ disagree with
the predictions of Ref.~\cite{DC} by a factor of $\approx2.5$. By
performing a linear extrapolation versus $1/\ln|t|$ they then find a
corrected result in reasonably good agreement. However, one might argue
that it is difficult to justify such a linear extrapolation, ignoring
$O(1/(\ln|t|)^2)$ terms, when the resultant correction is so large. In
any case, the analysis of Salas and Sokal\cite{SS} indicates that
finite-size effects cannot be excluded in the region that the
non-leading logarithmic corrections are small. The agreement with the
analytic predictions of Ref.~\cite{DC} may therefore be fortuitous,
particularly since, as we argue here, the latter may well be wrong.

Ziff and co-workers \cite{ziff}, following the appearance of Ref.~\cite{DC},
have reanalysed the percolation data which corresponds to the limit
$q\to1$. Previous quoted results for the ratio of mean cluster size
below and above $p_c$ ($\sigma_+/\sigma_-$, which is the $q\to1$ limit
of $\Gamma/\Gamma_L$, see Sec.~2) had ranged from 14 to 220. The
prediction of Ref.~\cite{DC}, based on simple extrapolation of the
results for $q=2,3,4$ (which appears to work to 10\% accuracy for
other amplitude ratios) gives a value around 74. However, Ziff et al.'s
value \cite{ziff} is $163\pm2$, in complete disagreement. (Note that in
percolation there are no finite-size effects and corrections to scaling
are generally small.) Ziff et al. have also measured the ratio
$\sigma'/\sigma_-$ of integrated correlations within the infinite cluster
to those in the finite clusters,\footnote{There appears to have been
some confusion over this in the literature in the past. For example,
Aharony and Stauffer\cite{SA} on p.~60 (2nd.~ed.) state that the mean
cluster size for $p>p_c$ is found by summing the connectedness function
$g(r)$ over $r$ and subtracting $P^2$, where $P$ is the probability of
a given site belonging to the infinite cluster. However, this would
give $\sigma_-+\sigma'$.} and find a value $1.5\pm0.2$, in perfect
agreement with our extrapolated value of $1.49$ in Eq.~\ref{149}.

What conclusion is to be drawn? There appears to be firm confirmation
of our new analytic results for the low-temperature ratio
$\Gamma_T/\Gamma_L$ both from $q=3$ and from percolation, while there
is strong evidence that the results of Ref.~\cite{DC} for the ratio
$\Gamma/\Gamma_L$ are incorrect. The most likely source of error lies
in the computation of the correlation function for $T>T_c$, since the
low-temperature calculations are verified by the other ratio.  We
recall that in fact all calculations in Ref.~\cite{DC} were performed
in the low-temperature phase, and that the order parameter form factors
for $T>T_c$ were inferred from those of the disorder operator for
$T<T_c$ by duality. In order to fix the ratio $\Gamma/\Gamma_L$ it is
therefore crucial to be able to fix the relative normalisation of the
order and disorder operator form factors. In Ref.~\cite{DC} this was
done assuming an extension (to the case of theories with internal symmetries)
of the factorisation result of \cite{DSC}. While this extension gives the
correct result for $q=2$, the analysis of this paper suggests that this 
may not be the case for $q=3,4$ \footnote{This would affect the predictions
of Ref.~\cite{DC} for the ratios $\Gamma/\Gamma_L$ and $R_C$, which are the 
only ones to be sensitive to the relative normalisation of order and 
disorder operators.}.

\section{Acknowledgements}

We thank Robert Ziff for discussions and for communicating his results
with us before publication. We have also benefited from correspondence
with M.~Caselle and A.~Sokal.
GTB gratefully acknowledges the High-performance computing group of
Utrecht University for computer time. JC's work was supported in part
by EPSRC Grant GR/J78327.

\end{document}